\documentclass[10pt,pra,onecolumn,superscriptaddress,showpacs,floatfix]{revtex4}

\usepackage{graphicx}
\usepackage{amssymb}
\usepackage{times}
\usepackage{amsmath}
\usepackage{amsthm}

\input epsf.tex
\usepackage{graphicx}
\usepackage{amsthm}

\begin{document}

\title{Weak value amplification via second-order correlated technique\thanks{This work was supported by the Union Research Centre of Advanced Spaceflight Technology (Grants No. USCAST2013-05) ,the National Natural Science Foundation of China (Grants No. 61170228£¬No. 61332019 and 61471239), and the Hi-Tech Research and Development Program of China (Grant No: 2013AA122901)}}% Force line breaks with\\

%\author{}\affiliation
% {}
\author{Ting Cui$^{\rm a)}$, Jing-Zheng Huang$^{\rm a)}$\thanks{Corresponding author. E-mail:jzhuang1983@sjtu.edu.cn}, Xiang Liu$^{\rm c)}$,  Gui-Hua Zeng$^{\rm a)b)}$ \\
\small
$^{\rm a)}${State Key Laboratory of Advanced Optical Communication Systems and Networks, }\\
\small
{Shanghai Key Laboratory on Navigation and Location-based Service,}\\
\small
{and Center of Quantum Information Sensing and Processing, Shanghai Jiao Tong University, Shanghai 200240, China}\\
\small
$^{\rm b)}${College of Information Science and Technology, Northwest University, Xi¡¯an 710127, Shaanxi, China}\\
\small
$^{\rm c)}${Shanghai Key Laboratory of Aerospace Intelligent Control Technology,}\\
\small
{Shanghai Institute of Spaceflight Control Technology, Shanghai 200233, China}}

\date{\today}

\begin{abstract}

We propose a new framework combining weak measurement and second-order correlated technique. The theoretical analysis shows that WVA experiment can also be implemented by a second-order correlated system. We then build two-dimensional second-order correlated function patterns for achieving higher amplification factor and discuss the signal-to-noise ratio influence. Several advantages can be obtained by our proposal. For instance, detectors with high resolution are not necessary. Moreover, detectors with low saturation intensity are available in WVA setup. Finally, type-one technical noise can be effectively suppressed.

\end{abstract}

\maketitle

\section{Introduction}\label{Introduction}

In the theory of quantum mechanics, the process of measuring an unknown state of a system by a meter can be described in two steps: coupling the system with the meter and measuring the meter to get information. For a standard projective measurement, where the coupling is strong, the outcome must be one of the eigenvalues of the system. On the other hand, when the coupling is weak, Aharonov, Albert and Vaidman\cite{1} showed that there is a very strange phenomenon that the outcome could be outside the range of the eigenvalues by means of appropriate pre-selection and post-selection on the system. This phenomenon is called weak value amplification (WVA), which plays a significant role in parameter estimation in recent years.\cite{2,3,4,5,6,7,8,9,10,11}

So far, many theories and experiments of WVA assume that the meter state is pure.\cite{1,12} However, in some cases, preparing a pure state can be difficult because decoherence is easily induced by environmental disturbance.\cite{13,14,15} In Ref.~\cite{16}, Young-Wook Cho, et al. considered this problem and proposed a new scheme applying incoherent meter state to decrease the experimental difficulty. They used pseudo-thermal light instead of laser to repeat the AAV experiment and confirmed the feasibility.

On the other hand, in the AAV weak measurement experiment with photon position as the meter, beam after post-selection is directly collected by a charge-coupled device (CCD).\cite{7,16} Though simple operation and calculation it has, larger displacement and better performance cannot be obtained for lack of sufficient data. It is natural to ask whether an  improvement can be achieved by using different detection techniques.

 Recently in Ref.~\cite{17}, Young's double-slit interference experiment was implemented by  second-order correlation detection, which was first proposed by Hanbury Brown and Twiss.\cite{18}  Researchers found that more data could be collected through two-dimensional second-order correlated function patterns and varied scanning methods, which effectively improved the imaging resolutions.

Inspired by this idea, in this paper, we propose a new scheme combining second-order correlated technique with pseudo-thermal light source to realize WVA. We find that WVA phenomenon can be observed from second-order correlated detection. By applying two-dimensional second-order correlation function patterns, we obtain different WVA amplification factors by means of varied scanning methods.

\section{Weak value amplification and second-order interference}
In the AAV weak measurement scheme,\cite{19} the interaction between a system and a meter can be described by a Hamiltonian:
\begin{eqnarray}
\hat H=\mathit{g}\hat A \otimes \hat P,
\end{eqnarray}
where $ \mathit{g} $ is the coupling constant, $ \hat A $ is the hermitian operator of the system, and $ \hat P $ is the momentum operator. The initial state, which is denoted as $ \vert i \rangle \vert \phi(x) \rangle $ , is converted to be
\begin{eqnarray}
&e^{-i\hat {H} t/ \hbar }\vert i \rangle\vert\phi(x)\rangle,
\end{eqnarray}
where $\vert\phi(x)\rangle$ is denoted as the meter, and $ \vert i \rangle $ is the pre-selected system state. Here we assume that the pre-selected state $ \vert i \rangle $ is the polarized state of a single photon in the horizontal-vertical basis, which can be denoted as $ \vert i \rangle=\cos{\alpha}\vert H \rangle + \sin{\alpha}\vert V \rangle $. Similarly, the post-selected state is $ \vert f \rangle=\cos{\beta}\vert H \rangle + \sin{\beta}\vert V \rangle $. After post-selection, the meter state evolves as
\begin{eqnarray}
\begin{split}
&\langle f \lvert e^{-i\hat {H} t/ \hbar } \rvert i \rangle \vert\phi(x)\rangle \\
&\approx(\langle f \rvert i \rangle-i \mathit{g} t \hat P \langle f \rvert \hat A \rvert i \rangle / \hbar)\vert\phi(x)\rangle \\
&\approx \langle f \rvert i \rangle (1-i\mathit{g} t \hat P A_w/ \hbar )\vert\phi(x)\rangle \\
&\approx \langle f \rvert i \rangle e^{i\mathit{g} t \hat P A_w/ \hbar}\vert\phi(x)\rangle.
\end{split}
\end{eqnarray}
where the high-order term of $\mathit{g}$ is neglected and $A_w=\frac{\langle f \lvert A \rvert i \rangle}{\langle f \vert i\rangle}$ is defined as the weak value.
%\begin{eqnarray}
% A_w=\frac{\langle f \lvert A \rvert i \rangle}{\langle f \vert i\rangle}.
%\end{eqnarray}

The feature of the weak value is that it can be arbitrarily large when the pre-selected state and post-selected state are nearly orthogonal.\cite{19} By choosing $ \alpha=\frac{\pi}{4} $, $ \beta=\alpha+\frac{\pi}{2}+\epsilon $, the weak value can be calculated to a simple form which equals
\begin{eqnarray}
\begin{split}
A_w &= \frac{\cos{\alpha}\cos{\beta}}{\cos{\alpha}\cos{\beta}+\sin{\alpha}\sin{\beta}} \\
&\approx \frac{1}{2} \cot{\epsilon},
\end{split}
\end{eqnarray}
where $\epsilon$ is a constant and can be infinitely small.

In our proposal, the experimental setup is shown in Fig.1. Here, we use the pseudo-thermal light generated by scattering a focused laser beam as the source. The beam is divided into two arms through a BS, which we denote the upper arm as the signal arm and the other one as the reference arm. In the signal arm, P1 and P2 are polarizers performing pre-selection and post-selection. The quartz plate Q performs a weak measurement. Detector D2 fixes on and only receives the photon in a fixed point. In the reference arm, D1 is a CCD to collect the photons in one dimension. This proposal is similar to ghost imaging.\cite{20}
\begin{figure}[!h]\center
\resizebox{11cm}{!}{
\includegraphics{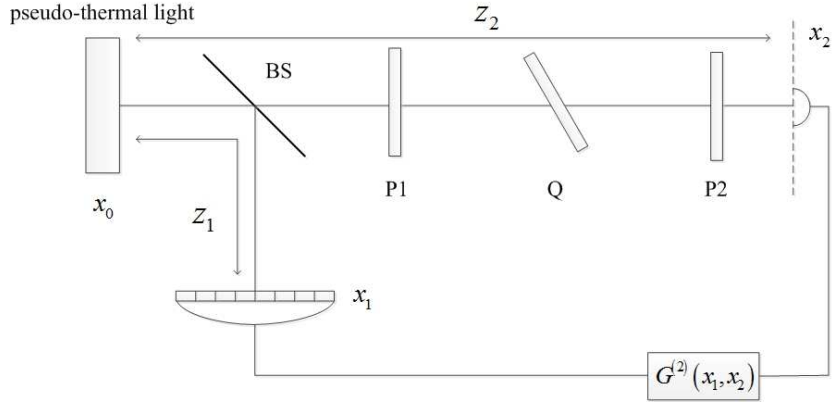}}
\caption{Experiment setup. The pseudo-thermal light is launched by scattering a focused laser beam. The light is divided into two arms by a BS. The signal arm (the upper path) is implemented by the AAV weak measurement system and the light is received by a single point detector. P1 and P2 are polarizers as pre- and post-selection. Q is a quartz plate. The light from the other arm is directly received by a CCD.}
\end{figure}

The optical field in the source plane is represented by $E(x)$. According to the optical system theory, the optical field in the detect plane is\cite{20}
\begin{eqnarray}
E(x_i)=\int E(x_0)h_i(x_i,x_0)dx, i=1,2,
\end{eqnarray}
where $ E(x_i),i=1,2 $ is the optical field in the reference and signal detector plane, respectively. And $ h_i(x_i,x_0),i=1,2 $ is the impulse response function of the reference arm and the signal arm, respectively. The second-order correlation function of any two points between two detector plane is defined as $G^{(2)}(x_1,x_2)$\cite{20,21}
\begin{eqnarray}
G^{(2)}(x_1,x_2)= \langle E(x_1)E(x_2)E^{*}(x_1)E^{*}(x_2) \rangle.
\end{eqnarray}

Substituting Eq.(5) into Eq.(6), we get\cite{20}
\begin{eqnarray}
\begin{split}
G^{(2)}(x_1,x_2)&=\iiiint h_1^{*}(x_1,x'_{01}) h_2^{*}(x_2,x'_{02}) h_1(x_1,x_{01})h_2(x_2,x_{02}) \times \\
&\langle  E(x'_{01})E(x'_{02})E(x_{01})E(x_{02}) \rangle \,dx'_{01}\,dx'_{02}\,dx_{01}\,dx_{02} \\
&=\langle I_1(x_1) \rangle \langle I_2(x_2) \rangle+ {\lvert G^{(1)}(x_1,x_2) \rvert}^2.
\end{split}
\end{eqnarray}
where $I(x_i),i=1,2$ is the intensity of the two detectors, and $G^{(1)}(x_1,x_2)$ is the first-order correlation function.

The second-order correlation function in Eq. (7) contains a background term $\langle I_1(x_1) \rangle \langle I_2(x_2) \rangle$ and a term which the information of our measurement can be extracted. Therefore, we calculate the correlation function of intensity fluctuations to obtain the information. \cite{20}
\begin{eqnarray}
\begin{split}
\langle \Delta I_1(x_1) \Delta I_2(x_2) \rangle &= {\lvert G^{(1)}(x_1,x_2) \rvert}^2 \\
&=\lvert \iint h_1^{*}(x_1,x_{01}) h_2(x_2,x_{02}) \langle E^{*}(x_{01})E(x_{02}) \rangle \,dx_{01}\,dx_{02} \rvert^2.
\end{split}
\end{eqnarray}

We assume that the pseudo-thermal light is completely incoherent and satisfies the Gaussian type. Therefore, the term of first-order correlation function is
\begin{eqnarray}
\langle E(x_{01})E(x_{02}) \rangle= \rm G_0 \exp(- \frac{x^2_{01}+x^2_{02}}{w^2_0})\delta (x_{01}-x_{02}),
\end{eqnarray}
where $\rm G_0$ is a constant and $w_0$ is the beam waist. $x_{01}$, $x_{02}$ are the points at any positions in the source plane.

In the signal arm, we assume that the two polarizers are thin enough that their thickness can be neglected. Consequently, both of the polarizers have no effect on the propagation of the light. After the weak-coupled process in the quartz plate, the emission of the beam contains two components, the ordinary and the extraordinary light. In the extraordinary light, the quartz plate will induce a distance $d$ due to birefringence and a phase difference $\theta$. So the impulse response function will be written into the sum of two terms. Therefore, according to the Fresnel-diffraction approximation, the impulse response function of the signal arm can be written as
\begin{eqnarray}
h_2(x_2,x_0)=\frac{e^{ikz_2}}{\sqrt{i\lambda z_2}}\lbrack \cos{\alpha}\cos{\beta} e^{i \frac{k}{2z_2}(x_2-x_0)^2} + \sin{\alpha}\sin{\beta} e^{i \frac{k}{2z_2}(x_2-d-x_0)^2} e^{i\phi} \rbrack,
\end{eqnarray}
where $\alpha$ and $\beta$  are angles of pre-selected state and pro-selected state, respectively. $k$ is the wave vector and $\lambda$ is the wavelength of the pseudo-thermal light, $z_2$ is the distance between the source and the detector D2.

In the reference arm, the light will directly propagate from the source to the detector D1. Thus, the impulse response function is
\begin{eqnarray}
h_1(x_1,x_0)=\frac{e^{ikz_1}}{\sqrt{i\lambda z_1}}e^{i \frac{k}{2z_1}(x_1-x_0)^2},
\end{eqnarray}
where $z_1$  is the distance between the source and the detector D1.

Substitute Eq.(9), Eq.(10) and Eq.(11) into Eq.(8), the correlation function of intensity fluctuations is
\begin{eqnarray}
\begin{split}
\langle \Delta I_1(x_1) \Delta I_2(x_2) \rangle \propto \lvert \tilde{I}(\frac{x_1-x_2}{\lambda z}) \rvert^2 \cos^2{\alpha} \cos^2{\beta} + \lvert \tilde{I}(\frac{x_1-d-x_2}{\lambda z}) \rvert^2 \sin^2{\alpha} \sin^2{\beta} \\
+ 2 \lvert \tilde{I}(\frac{x_1-x_2}{\lambda z}) \rvert \lvert \tilde{I}(\frac{x_1-d-x_2}{\lambda z}) \rvert \cos{\alpha}\cos{\beta}\sin{\alpha}\sin{\beta}\cos(\frac{\pi}{\lambda z}d(d-x_2)+\phi),
\end{split}
\end{eqnarray}
where $\tilde{I}(\frac{x_1}{\lambda z})$ is the Fourier transform of $I(x_0)$ under the condition of $z_1=z_2=z$.

The position of the detector on the signal arm is fixed. And, we set this fixed position at the center of the signal plane. Thus the intensity fluctuation is
\begin{eqnarray}
\begin{split}
\langle \Delta I_1(x_1) \Delta I_2(x_2 =0) \rangle \propto \lvert \tilde{I}(\frac{x_1}{\lambda z}) \rvert^2 \cos^2{\alpha} \cos^2{\beta} + \lvert \tilde{I}(\frac{x_1-d}{\lambda z}) \rvert^2 \sin^2{\alpha} \sin^2{\beta} \\
+ 2 \lvert \tilde{I}(\frac{x_1}{\lambda z}) \rvert \lvert \tilde{I}(\frac{x_1-d}{\lambda z}) \rvert \cos{\alpha}\cos{\beta}\sin{\alpha}\sin{\beta}\cos(\frac{\pi}{\lambda z}d^2+\phi),
\end{split}
\end{eqnarray}
For a small $d$, from setting $\alpha=\frac{\pi}{4}$, $\beta=\alpha +\frac{\pi}{4}+\epsilon $, $\phi=2\pi$, Eq.(13) can be simplified by
\begin{eqnarray}
\langle \Delta I_1(x_1) \Delta I_2(x_2 =0) \rangle \propto \exp(2(\frac{x_1-dA_w}{w_1})^2),
\end{eqnarray}
where  $A_w$ is the weak value and $w_1=\frac{\lambda z}{\pi w_0}$.

From Eq.(14), we can see clearly that there is a transverse position shift from the imaging beam. And the amplification of this AAV system is exactly $A_w$. This result suits the traditional AAV weak measurement in Ref.~\cite{19}, but only the radius of the beam waist is changed. Thus, using intensity fluctuation correlation can recover the imaging as well and extract the displacement which we mostly need in weak measurement system.

In fact, from Eq.(12) we can know that the second-order correlation function is the correlation of any two points in two detected planes, respectively. If we choose different scanning methods of the two detectors,\cite{17} we can get different transverse position displacements. And that will not be seen in one-path system.

We replace the one-point detector of reference arm in FIG.1 with another CCD2 completely the same as CCD1 in signal arm. And we construct a two-dimensional second-order correlation function with $x_1$ and $x_2$ being the horizontal and vertical coordinates. The pattern is shown in FIG.2.

By choosing different lines on the two-dimensional second-order correlation function figure, any displacements of AAV weak measurement will be obtained. If we rotate the scanning line along different direction clockwise, we can get larger displacements and amplifications than that in the traditional weak measurement. For example, line (b) describes the case where we just proposed above. Notice that, scanning the detectors through this line cannot get the largest displacement and amplification. Also, the curve is shown in FIG.3 (b). Line (a) describes the case where we scan the two detectors in opposite direction. And the result is shown in FIG.3 (a). Line (c) shows a scanning method with $x_1=x$ and $x_2=0.5x$. We obtain a larger displacement and amplification, shown in FIG.3(c). We change the scanning method to line (d) with $x_1=x$ and $x_2=0.75x$, and the result is in FIG.3 (d). From the statements above, we can draw a conclusion that the scanning method of the second-order correlation function figure has different effect on weak value amplification. And if we choose a method properly, better weak value amplification will be seen.
\begin{figure}[!h]\center
\resizebox{11cm}{!}{
\includegraphics{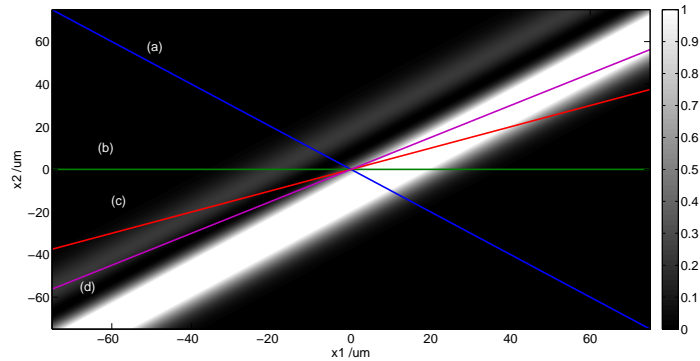}}
\caption{Two-dimensional second-order correlation function in Eq.(12) with $x_1$ and $x_2$ being the horizontal and vertical coordinates. The four lines illustrate different methods of scanning the detectors: line (a) for $x_1=x$ and $x_2=-x$; line (b) for $x_1=x$ and $x_2=0$ ; line (c) for $x_1=x$ and $x_2=0.5$ ; line (d) for $x_1=x$ and $x_2=0.75x$.}
\end{figure}
\begin{figure}[!h]\center
\resizebox{11cm}{!}{
\includegraphics{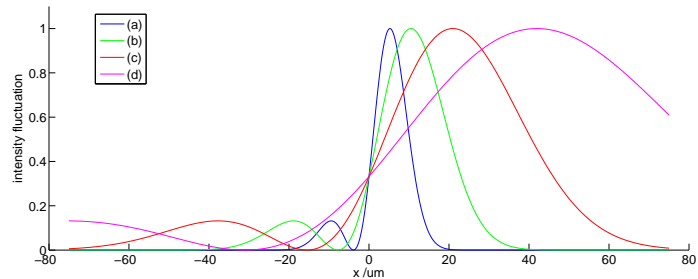}}
\caption{Cross-sectional curves of (a), (b), (c) and (d) represent the corresponding scanning lines in FIG.2.}
\end{figure}

\section{Discussion}
From FIG.2. and FIG.3., it is clear that using different methods of scanning, arbitrary amplifications within the range of the size of CCD can be obtained. To prove the performance, we use the simulation data completely identical with the traditional AAV weak measurement experiment.\cite{19}. As proposed before, line (b) is the traditional method. We calculate the amplification and the result is the same as the experiment proposed in.Ref.~\cite{19} In fact, from Eq.(15), it is shown that the displacement or the amplification has no relationship with how we detect. Therefore, using line (b) as the basic line, we rotate it clockwise or counterclockwise and get smaller or larger amplification. These results are shown in FIG.3.
From FIG.3, it is also shown that the beam waist is changed as well as amplification. The variation of amplification and beam waist may also influence the signal-to-noise ratio (SNR) which has been discussed in Ref.~\cite{22,23} and minimum resolvable signal in Ref.~\cite{23,24}. Here, the SNR can be written as
\begin{eqnarray}
SNR=\frac{A_w d}{\sigma / \sqrt{N}},
\end{eqnarray}
where $N$ is the average photons after post-selection and $\sigma$ is the standard deviation of the intensity fluctuation curve. And the minimum resolvable signal $d_{min}$ is denoted as
\begin{eqnarray}
d_{min}=\frac{\sigma}{ A_w \sqrt{N}},
\end{eqnarray}
Since the post-selection is fixed in our setup and the scanning methods do not change the average photons, $N$ remains constant. Because the analytical solution is hard to derive, we make a numerical simulation of $\sigma$ and $A_w$ in each scanning method. The variation tendency between $\sigma$ and $A_w$ is shown in Table 1.

\begin{center}
{\footnotesize{\bf Table 1.} Numerical Simulation of $\sigma$ and $A_w$ in each scanning method.\\
\vspace{2mm}
\begin{tabular}{cccc}
\hline
scanning method   & $A_w$/um            & $\sigma$/um   & $A_w/\sigma$ \\\hline
$a$              & 5.18                 &  8.41        & 0.6161\\
$b$               & 10.44                & 16.82         & 0.6205\\
$c$               & 20.95                & 34.23         & 0.6118\\
$d$               & 41.97                & 68.47         & 0.6129\\
\hline
\end{tabular}}
\end{center}

We can see clearly that the ratio between $A_w$ and $\sigma$ varies in a very small range, and thus we can  conclude that both SNR and $d_{min}$ are independent of scanning methods.

Though, the SNR and resolution cannot be improved whatever scanning methods we choose in our proposal. The advantages are remarkable. First, large displacement will decrease the resolution requirement for detectors. Besides, due to post-selection probability, detectors with low saturation intensity is allowed to use in WVA setup.\cite{24} Finally, it can also reduce type-one technical noise which has been investigated before.\cite{24}

\section{Conclusion}
In this paper, we propose a new scheme to implement AAV weak measurement. The theoretical analysis manifests that WVA experiment can also be realized through second-order correlated technique. Moreover, we set up two-dimensional second-order correlated function pattern and describe how the amplification can be improved by choosing different scanning lines. Our theoretical model will pave the way to apply WVA setup in second-order correlated systems, such as ghost imaging. Furthermore, high amplification can reduce technical requirement for devices which lowers the cost and decreases experiment difficulty. Finally, WVA can effectively reduce technical noise. We believe our theory can have a bright prospect for weak measurement and second-order correlated technique.

\end{document}